\documentstyle[sprocl]{article}

\def\be{\begin{equation}}
\def\ee{\end{equation}}
\def\bea{\begin{eqnarray}}
\def\eea{\end{eqnarray}}

\begin{document}

\title{Black holes, string theory and quantum
coherence~\footnote{The content of
this note, even if with slightly different wording, was sent to the
archives as hep-th 9706157.  Several discussions since then, specially
in occasion of the Spinoza Meeting on Quantum Black Holes (Utrecht 29 June
-  3 July 1998) have shown the convenience of its publication} }

\author{Daniele Amati}

\address{International School for Advanced Studies (SISSA) - Trieste -
Italy\\
INFN - Sezione di Trieste\\ email: amati@sissa.it}

%

\maketitle
\abstracts{
On the basis of recently discovered connections between D-branes and black
holes, I show how the information puzzle is solved by superstring theory
as the fundamental theory of quantum gravity.  The picture that emerges is
that a well-defined quantum state does not give rise to a black hole even
if the apparent distribution of energy, momenta, charges, etc. would
predict one on classical grounds.  Indeed, geometry - general relativistic
space time description - is unwarranted at the quantum microstate level.
It is the
decoherence leading to macrostates (average over degenerate microstates)
that provides - on the same token - the loss of quantum coherence, the
emergence of a space time description with causal properties and, thus,
the formation of a black hole and its Hawking evaporation.}

\vskip1cm

  Many extraordinary coincidences between string 
theory and black hole physics have been recently uncovered and 
different opinions have been advanced on how these coincidences 
may ``explain`` or ``solve`` the well-known information loss 
puzzle.  The idea$^{(1)}$ that very massive 
string excitations should represent black holes has been better 
substantiated by analysing massive BPS states that, as known, 
have properties that are not renormalized, i.e. do not depend 
on the coupling strength.  

  In the weak string coupling (g) regime, D-branes in four$^{(2)}$ 
and five$^{(3)}$ dimensions with a convenient number of charges 
have been studied.  BPS states  have been counted as well as 
nearly BPS states for certain regions of moduli space where perturbative 
computations are feasible$^{(4)}$.  Decay rates have also been 
computed$^{(5) - }$by averaging over the many initial states 
- and shown to have, a typical thermal distribution. The moduli  
independence of these results allow the conjecture$^{(6)}$
of their validity beyond the moduli region where they were computed. 
 And their g independence (also suggested by non-renormalization 
arguments$^{(7)}$) may imply that they could be continued beyond 
the weak coupling regime.

  An independent treatment - on totally different grounds - of 
the strong coupling regime substantiates that impression.  The 
large coupling description of the 4 and 5 dimensional systems 
just discussed is found by solving the 10-d supergravity equations 
after reduction on the same compact manifold used for the D-brane description.  
The solution generates a metric$^{(8)}$ that depends on parameters 
that are related to the charges through the moduli of the compact 
manifold.  The metric shows an event horizon even in the extreme 
limit;  its area in this limit gives the Beckenstein-Hawking 
entropy of extremal b.h.  This entropy and the ADM mass 
coincide exactly with the mass and entropy (given by the log 
of the state multiplicity) of the BPS state with the same charges 
as computed from D-branes in the small coupling regime.

  For nearly extremal b.h. the entropy, the rate and the spectrum 
of evapora-tion$^{(9)}$ - obtained by solving wave equations in 
the corresponding metric background - coincide again$^{(5)}$ 
with those computed for small g.  And, even more remarkably, 
also deviations from black body spectrum agree$^{(10)}$.  These 
magic coincidences between such different calculations gave confidence 
to the g$^{ }$ continuation between a unitary D-brane description 
and the conventional black hole with its information loss.  

  Different interpretations of this apparent contradiction have 
been advanced. Hawking questioned the D-brane formalism because 
the causal properties of space time are not properly taken into 
account.   And even more so having shown$^{(11)}$ that the inconsistency 
cannot be assigned to the lack of identificability of D-brane 
states due to their prompt decay.  String theorists$^{(12)}$ 
favour the attitude that black body radiation is only an approximation 
and that the connection with unitary D-brane formalism guarantees 
information retrieval in b.h. evaporation.  This even more so 
since duality indicates situations that have flat unitary realizations 
for g$<$$<$1 as well as for g$>$$>$1 with a b.h. intermediate 
region with quite different space time geometry but with an alledgedly 
common spectrum.

  Let me share the consensus of a smooth g behaviour by considering 
an S-matrix approach where a continuation has a clear meaning 
and let me start in the small coupling regime where a perturbative string 
approach is granted.

  If over a well-defined BPS state impinges from far away - where 
for small g the space is flat - a well-defined quantum state 
(say a graviton) the S matrix, calculable by perturbative string 
theory, will be unitary.  It describes the absorption of the 
graviton generating open strings on the D-brane then decaying, 
through an arbitrary number of steps, to some BPS state plus 
outgoing closed string ground states (as gravitons and scalars).

  The physical content of this large set of S matrix elements is better
analysed by computing final state correlation functions that give
semi-inclusive quantities as multiplicities, spectra or multiparticle
correlations.  Old string theorists will remember the techniques used to
directly compute these correlators from multiparticle amplitudes, order
by order in g.  They will also recall, however, the need to recurr to the
full set of
final state correlators in order to disentangle the high degeneracy of
initial states.  This means that if we change the choice of the initial
BPS state (among the very many degenerate ones) or if we consider two
gravitons impinging instead of a single one, or if we change g, all these
correlators will be modified in a non-trivial and coherent way.  The whole
set of correlators represent the complete memory of the original state.

  The result we discussed before concerning the Hawking spectrum,
tells us that if we now perform an average over the very many possible
initial (degenerate) states, all multiparticle correlators will average to
zero apart from the single particle spectrum that - by energy conservation
- averages to a thermal one.  And this for each g, thus order by order in
the string loop expansion.  This is perhaps not too surprising:
the average
over the very many degenerate microstates washes out all information over
the initial identity leading to the informationless black body
radiation. Let me stress that this average over degenerate microstates is 
implied in any classical limit.

  In increasing g towards a black hole regime, it is the quantum S matrix
that should be analytically continued.  Its analytic structure may be very
complex, with new singularities being eventually formed, but with
unitarity preserved with the concurrence - as before - of the very many
phases that depend on the initial (black holish) microstate.  These are
essential in order to determine the many non-trivial correlation
functions.  The fact that these averaged to zero independently from g for
small g - giving the same Hawking spectrum of the large g regime - shows
that no
new physics (new singularity in the g continuation) has to be invoked in
order to understand the spectrum and entropy of macroscopic black holes
(statistical collection of microscopic states).

  This same reasoning, however, shows that the decay spectrum of a single
microstate differs crucially from a
black body one.  This was the case for all small g, thus for all g by
continuation in a context in which no new physics is evoked.  We thus
expect for a
"would be black holish" microstate, far from vanishing correlators that
encode the whole information about the (coherent) formation process.

  It is apparent that a single microstate - even in the large g regime -
has not much to do with a black hole and that it is only the decoherence
implied by the macroscopic description (i.e. average over microscopic
states) that generates the black hole physics.  Such a statement needs,
however, a parallel understanding of why it is only at the macroscopic
level that the geometrical interpretation of general relativity emerges
with its causal properties, singularities and event horizons.  

  Superstring theory contains gravity in the infrared limit; for
frequencies much smaller than the string scale, the
Einstein classical equations appear as the non renormalization
($\beta$=0) condition.  At the quantum level,
however, fluctuations at the string scale will generate all other (massive)
background fields
which will appear (thanks to the $\beta$=0 condition) in a large system of
coupled equations together with the metric field.  In a more common
language, this implies a very large number of quantum hairs. These many
non-metric coupled fields, that inhibit a geometrical space time
description, are expected to have quickly varying phases so as to be
averaged out in the decoherence procedure implied by classical (or
mesoscopic) physics.  This is perhaps not surprising, superstrings are
pregeometric quantum theories in which even a space time
description is not warranted: X$\mu$ are operators and it is only at
the mesoscopic level (in which quantum string fluctuations are averaged
away) that they appear as coordinates parametrizing a metric
space.  It is
thus this decoherence that generates a geometrical space time
description (i.e. general relativity) and with it, causal properties event
horizons and the paraphernalia of black hole physics.

  The consideration up to now of charged extremal or near extremal black
holes or, in the string language, solitonic D branes which are BPS or
quasi BPS states, was an essential step in order to identify and count
stable or nearly stable states.  This allowed the consideration of S
matrices with well-defined quantum initial states, excited in the process
and evaporating back to stable final states.

  For unstable states (high string excitations or Schwarzschild b.h.) a
consistent quantum treatment has to comprehend both formation and decay.
In the small g regime this is anyhow the conventional approach of
perturbative string amplitudes.  In the b.h. regime it implies the
consideration at a consistent quantum level of both the formation and
evaporation of a b.h, thus avoiding, ultimately, the hybrid theoretically  
procedure of quantizing in the presence of a purely classical
solution.  It is obvious that one might unambiguously prepare imploding
states (spherical waves or even high energy low mass particles colliding
at very small impact parameter) at large separations where a flat metric
is granted.  At a classical level and even at a semiclassical one
$^{(13)}$ (i.e. with radiation) black holes would be formed in these
conditions losing memory of the state they originated from. 

  This is not
the case at the quantum level. Again, at small g where everything is in
principle calculable, final state correlators are far from vanishing
and contain all the information needed to disentangle the initial state
(unitarity). No new physics has to be invoked in order to continue to
the large g black hole regime: it is the presence of non metric fields of
arbitrary high tensorial rank that
avoids
the Schwarzschild singularity of the usual general relativistic (metric)
solution. And, again, it is the mesoscopic decoherence implied by
averaging over microstates, that on one hand averages out the
correlators that carry all the microscopic information thus leading to a
black body (or approximate black body) spectrum and, on the other hand,
washes away the non metric fields thus leading to a geometric space time general
relativistic picture with causal properties, horizons, black holes, etc.

  At this level it may be sound to ask if this decoherence may be avoided
in a gedanken experiment so as to show the full quantum structure of the
fundamental gravity theory.  Could, for instance, a classically
expected black hole be avoided by preparing well-defined coherent
imploding states?  In my opinion the infrared properties of gravitation
may
jeopardize this possibility.  Indeed, it seems even conceptually hard to
avoid incoherent arbitrary soft gravitons and, with them, their high
decoherence power due to the very large density of microstates.

  Let me stress the important role that the high degeneracy of
states had in the smooth merging of a
unitary microstate description into a black hole macrostate one.
Qualitatively, it is apparent that a degeneracy that grows exponentially
with the mass is a border-line between a tendency of states, interacting
through vertex operators, to split as in particle physics or join as
for b.h. due to the final states multiplicity.  One would be tempted to
think that consistent fundamental theories of quantum gravity have to have
such a degeneracy in order to lead to macroscopic general relativity.  No
wonder, in this sense, that superstring theory is a good candidate while
supergravity is not.  It would be interesting to understand how other
proposals that have been advanced, such as topological gravity, may solve
this  problem in their attempt to qualify as possible consistent theories 
of quantum gravity.

This work was partially supported by EC Contract no.
ERBFMRXCT960090 and
by the research grant on Theoretical Physics of Fundamental Interaction of
the Italian Ministry for
Universities and Scientific and Tecnological Research.


\vskip2cm

\end{document}